# Application of Neural Network Algorithm in Propylene Distillation

Jinwei Lu, Ningrui Zhao

School of Chemical Engineering, East China University of Science and Technology, Shanghai, 200237

**Abstract:** Artificial neural network modeling does not need to consider the mechanism. It can map the implicit relationship between input and output and predict the performance of the system well. At the same time, it has the advantages of self-learning ability and high fault tolerance. The gas-liquid two phases in the rectification tower conduct interphase heat and mass transfer through countercurrent contact. The functional relationship between the product concentration at the top and bottom of the tower and the process parameters is extremely complex. The functional relationship can be accurately controlled by artificial neural network algorithms. The key components of the propylene distillation tower are the propane concentration at the top of the tower and the propylene concentration at the bottom of the tower. Accurate measurement of them plays a key role in increasing propylene yield in ethylene production enterprises. This article mainly introduces the neural network model and its application in the propylene distillation tower.

## 1 Introduction

The neural network control system has good intelligence and robustness. It can handle the control problems of complex industrial production processes with high dimensions, nonlinearity, strong coupling and uncertainty. Its notable feature is its self-learning ability, continuously correcting the connection weights between neurons, and discretely storing them in the connection network. So it has good mapping capabilities for nonlinear systems and systems that are difficult to model [1].Therefore, it can be used in the multi-parameter optimization process [2] such as chemical engineering process and control.

The propylene distillation tower plays a decisive role in the quality of the finished propylene.

The double-tower depropanizing system is the pre-system of propylene distillation. The rectification effect directly affects the propylene distillation process. If there is no corresponding automatic control to cooperate with it, it will be difficult to achieve the desired effect [3].

The non-linear dynamic phenomenon of the distillation column is very severe. If the feed flow rate is increased or decreased by a certain amount from the steady state, When the system reaches a new equilibrium, the time constant and gain of the transition process response are very different. Factors such as the difference in the rate of heat release , heat absorption and the irreversibility of the chemical reaction process, lead to this consequence. In the classical process control theory, this kind of problem has not been solved well. As a nonlinear expression, neural networks can be used to establish the input and output models of the system. They can be used as the forward or reverse dynamic model of the controlled object, or the approximation model of the controller. They also can be used to describe the performance evaluation device. This article briefly describes the neural network algorithm and its research progress, and summarizes the application of neural network algorithm in propylene distillation.

## 2 Neural network algorithm and its research progress

Artificial neural network ( ANN) [4] is an abstraction and modeling of the human brain or biological neural network. It has the ability to learn from the environment and adapt to the environment in a biological-like interactive manner. The definition of artificial neural network defined by Hecht Nielsen [5] is: "An artificial neural network is a computer system formed by a number of very simple processing units connected to each other in a certain way. The system processes information by its dynamic response to external input information."

The development history of neural networks is divided into four stages: rise, depression, prosperity and climax. In 1943, McCulloch [6] published an article on neural networks, marking the beginning of the era of neural network scientific research. Later Minsky and Papert [7] pointed out some problems in neural networks, which caused its development to almost stagnate. Until the 1980s, Professor Hopfield proposed the Hopfield [8] network model, which brought the development of neural networks out of the trough. After that, there was a period of rapid development of neural networks.

In recent years, neural networks have achieve great development. Researchers have proposed hundreds of different neural network models based on actual problems. These models can basically be divided into three categories: feedforward, feedback and self-organization. Different types of networks address different problems. The most important feedforward type is BP network [9] and RBF network [10], the most important feedback type is Hopfield network [11], and self-organizing network is mainly ART network [12]. Recently, fractal neural networks [13] and evolutionary neural networks [14] have appeared.

**2.1 Artificial neural network model**

An artificial neural network is a system that uses a physically achievable system to imitate the structure of the human brain neural network. Also, the network have functions of certain memory, calculation and judgment. Artificial neural network comes from the simulation of the actual neural network of the human brain. Let's first briefly analyze the structure and function of biological neural tissue.

Nerve cells are called neurons for short. The part of the cell nucleus is the cell body. Many nerve fibers extend from the cell body. The longest one becomes an axon, and its terminal turns into many small branches. Small branches become nerve endings. Branches starting from other branches that come from the cell body become dendrites. A cell communicates with the dendrites of other cells through axons. Therefore, dendrites are the input of cells and axons are the output of cells. The contact interface between nerve terminals and dendrites is called synapse.

From the perspective of the structure of the human brain, it is composed of a large number of nerve cells [15]. A neuron has corresponding input channels and output channels connected to it. In terms of function, the cell body can be regarded as a primary signal processor. Each cell performs a certain basic function, such as excitation and inhibition. When a signal is transmitted from a neuron to another cell body via a synapse, two effects can be produced: the potential of the cell receiving the signal increases or decreases. When the potential in the cell body is lower than a certain threshold, the signal is excited. In this moment, the cell is called an excited state [16]. When the potential in the cell body is lower than a certain threshold, no signal input is generated. It is in a state of inhibition.

For different downstream neurons, the potential changes caused by the signal are different. Different neurons have different intensities of action or connection strength. The process that a neuron accumulates and sums up excitatory (stimulating) or inhibitory signals from different dendrites is called integration.

In the artificial neural network model, the neuron is the most basic processing unit. It has the basic characteristics of biological neural organization. The neural network is a complex network system composed of a large number of simple neurons. Therefore it is widely used in many fields such as speech processing[17], artificial intelligence recognition[18], vehicle tracking[19], signal processing[20], stock market index prediction[21], medical diagnosis[22].

## 2.2 Basic principles of BP neural network

The feedforward BP algorithm was proposed by Rumelhart, Hinton and Williams [23]. It consists of a large number of nodes and their interconnections, including input layer, hidden layer and output layer. It is a multi-layer feedforward neural network trained according to the back propagation algorithm [24]. The basic idea is: the learning process consists of the forward propagation of data and the back propagation of errors. When the signal is propagating forward, the input data enters the model through the input layer without any processing at first. Then, it undergoes the first processing in the hidden layer. Finally, it reaches the output layer for the second processing. At this time, if the error between the network output and the actual output does not meet the accuracy requirements, the error will be backpropagated. The back propagation of errors is to propagate the difference between the network output and the actual output through the hidden layer to the input layer. In this process, the error signal of each processing unit is obtained, and the weight of each unit is modified by this signal. Until the error between the network output and the actual output is within the accuracy, or the network training reaches the set number of learning times, the network training is terminated. The modeling comes into an end [25].

## 2.3 Basic principles of RBF neural network

RBF neural network is a forward network with a single hidden layer. It can not only be used for function approximation, but also prediction[26]. In theory, it can approximate any nonlinear function [27]. It consists of an input layer, an implicit radial base layer and an output linear layer. Its network structure is shown in Figure 2. The radial basis function [28] is radially symmetric, and the Gaussian function is shown in equation (11).

$$R_i(x) = \exp\left(-\frac{\|x - c_i\|}{2\sigma_i^2}\right), i = 1, 2, \cdots, p \tag{11}$$

Among them:

X is the m-dimensional input vector, $c_i$ is the center of the i-th basis function, $\sigma_i$ is the variance of the i-th basis function, p is the number of perceptual units.

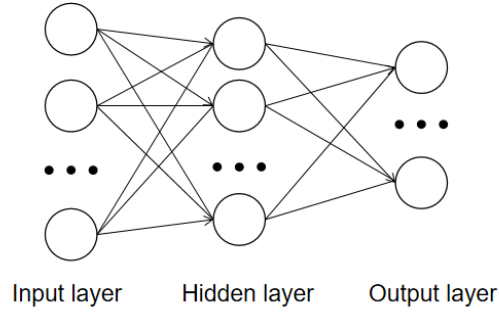

**Figure 2 RBF network structure diagram**

The input layer of the RBF network realizes the nonlinear mapping from x to $R_i(x)$ [29], and the input layer realizes the linear mapping from $R_i(x)$ to $y_R$. The equation of $y_R$ is shown in equation (12).

$$y_R = \sum_{i=1}^{p} w_{ki} R_i(x), k = 1, 2, \cdots, q \tag{12}$$

where $q$ is the number of output nodes, $w_{ki}$ is the adjustment weight between the *k*-th output layer and the *i*-th hidden layer.

## 2.5 RBF neural network algorithm description

(1) Determination of the input layer node. In the process of propylene rectification, a variety of process parameters affect the quality of the product. The main process parameters of the tower reactor temperature, tower pressure and sensitive plate temperature are selected as the input nodes of the neural network.

(2) Determination of the output layer node. The tower top composition is used as the evaluation index in this network.

(3) Determination of hidden layer nodes. In the RBF neural network, the hidden layer plays a key role. The number of hidden layers affects the accuracy, error, and speed of the model. So it is necessary to make tentative choices based on actual conditions and then gradually optimize.

(4) Standardized processing. Because the value range and dimension of each parameter are very different, the influence of each parameter on the surface roughness cannot be determined under the same standard. In order to make the neural network converge faster and have better generalization ability during the training process, it is necessary to standardize the sample data first and convert all the data into numbers between [0,1]. Take the maximum and minimum normalization method , the function form is shown in equation (13).

$$X_k = \frac{x - x_{\min}}{x_{\max} - x_{\min}} \tag{13}$$

$X_k$ is the normalized data, $x_{\min}$ is the smallest number in the data, $x_{\max}$ is the largest number in the data.

(5) Training of the network model [30]. The first 15 sets of data are used as training samples, and 16-20 sets of data are used as test data. It is necessary to create an RBF neural network with the newrbe function in the neural network toolbox that comes with matlab. In the RBF neural network, the number of input layer nodes is 3, and the output layer nodes are 4. The number of hidden layer nodes is automatically set by the newrbe function. Spread is the expansion coefficient of the radial basis function. If the spread is too large or too small, it is not conducive to the prediction of the neural network. If the spread is too small, the input vector may not be able to cover the area, and the number of neurons needed will increase. If it is too large, it will overlap areas between neurons, causing overfitting [31] and the increase of numerical calculations difficulty. Therefore, the trial and error method is used to determine the value of spread. Respectively 15, 19, 20, 21 and 25 are used for prediction. After comparative analysis, the spread is determined to be 20.

## 3 Simulation and optimization control research of propylene distillation tower

Since the industrialization of petroleum hydrocarbon cracking to ethylene [32], the petrochemical industry has developed rapidly. The so-called petroleum hydrolysis refers to the use of petroleum hydrocarbons as raw materials to obtain ethylene through high-temperature cracking, compression, and separation. At the same time, propylene, butadiene, benzene and other by-products can also be obtained [33]. The propylene distillation tower is the last link of the propylene product. The distillation process plays a decisive role in the quality of the propylene product. The traditional control scheme cannot achieve better gas-liquid exchange matching, so that the product propylene composition changes greatly. To achieve advanced control of the quality of propylene products, we must first establish a soft-sensing prediction model of propylene product quality[34] to realize online estimation of the quality index. Then, we must formulate a corresponding control plan based on the process mechanism to develop a suitable device Advanced control system. Because the propane product of the tower reactor is the raw material of the ethylene cracking furnace. The high content of propylene not only causes the waste of propylene resources, but also causes the coking rate of the ethane cracking furnace to be fast. Therefore, the column propane product[35] mainly uses the amount of propylene contained in it as the quality index.

Propane ($C_3H_8$) occupies an absolute proportion of the impurities in the propylene product, which directly affects the purity of the product propylene. So $C_3H_8$ (ppm) is selected as the dominant variable. Propylene rectification tower is a complex device. And product quality of the tower is not only affected by the operating conditions of the tower itself, but also by the fluctuations in the operation of each section above. The propylene distillation system [36] separates propylene and propane, which have close boiling points and relatively low volatility. Therefore, the separation and the process operation is difficult to achieve. From the perspective of process mechanism, we qualitatively analyze the main factors affecting propylene.

For the same tower kettle heating capacity, changes in the feed (F1C472) load will cause changes in the tower temperature, pressure and tower pressure difference. Therefore, the separation effect will be affected. However, because the tower has a diameter of 3.8 meters and has 153 trays, the material composition change in the top of the tower caused by the fluctuation of the feed material has a large lag.

For the operation in the tower, the gas-liquid exchange must meet a certain matching ratio. If the reflux rate of the rectification tower is large and too much liquid flows down in the tower, the product purity can be improved. But this is at the cost of increasing the tower reboiler load and condenser energy consumption. And because the propylene distillation tower bottom reboiler is heated by about 88"C quench water, the quench water is directly affected by the operating conditions of the cracking furnace. The quench water flow and temperature fluctuate greatly. The heating capacity is often insufficient. Although energy consumption is saved, it may cause the product to be unqualified if the reflux is too small. According to statistical analysis, the ratio of reflux to discharge is closely related to the propylene concentration [37];

1) The temperature at the top of the tower is a characteristic of the top components when the tower pressure is constant. As the overhead material is very pure, it is basically propylene. The fluctuation of the content of the impurity propane causes small fluctuations in the top temperature. But it is very sensitive.

2) The temperature of the 136 plates in the tower is the sensitive plate temperature. For binary components, the temperature change directly reflects the separation effect in the tower when the pressure is relatively stable. And it indirectly reflects the product concentration index of the bottom of the tower.

3) The change of tower pressure reflects the distribution of vapor and liquid in the tower and characterizes the separation effect in the tower.

4) The temperature of the tower kettle reflects the operating conditions of the tower kettle. And it is closely related to the tower top and the product concentration of the tower kettle.

## 4 Application of neural network algorithm in propylene distillation tower

The neural network control system has good intelligence and robustness. It can deal with the control problems of complex industrial production processes with high dimensions, nonlinearity, strong coupling and uncertainty [38]. Its notable feature is the ability to learn. It can continuously modify the connection weights between neurons, and discretely store them in the connection network. So it has good mapping capabilities for nonlinear systems and systems that are difficult to model [39]. Soft measurement technology has been widely used in process control systems. With the development of production technology and the increasing complexity of production processes, it is necessary to control important process variables of the system optimally in real time in order to ensure the safer and more efficient operation of production devices [40].

The propylene distillation tower is the key unit of the ethylene production plant. Its main function is to separate the binary mixture of propylene and propane into the polymerization grade propylene produced at the top of the tower and the propane product at the bottom of the tower. The design quality index of the propylene product at the top of the tower is that the concentration is greater than 99.6 in percent (polymerization grade). At the same time, the propylene concentration in the bottom of the tower generally does not exceed 5.0 in percent. Propane concentration and propylene component analyzers often become unstable. In order to ensure the purity of polymerization-grade propylene at the top of the tower, operators often set a larger reflux ratio. So the top quality is excessive and the loss of propylene at the bottom of the tower is increased. Therefore, it is necessary to start from the mechanism of propylene rectification and estimate the concentration of propane at the top of the tower and the concentration of propylene at the bottom of the tower online in real time.

Yuan RH[41] established a 4-layer feedforward neural network structure in order to predict the composition of the bottom product of the distillation tower as a forward model of the dynamic system. He also used the BP learning algorithm to train the neural network. The neural network inverse model of the dynamic system is established as the controller of the system. The neural network internal model control structure is used to control the bottom product composition according to the second stage temperature of the distillation tower. According to the experiments, compared with gas chromatography, the neural network method can approximate any non-linear mapping with arbitrary accuracy, provide product estimates faster, and enable the control system to take more timely measures to improve the control effect. Using the neural network internal model control structure, the neural network estimated value is used in the control. The actual process output and the expected deviation can be eliminated by the control method to eliminate the estimation error. To realize the static error inference control, he calculated the estimated error and used it to correct the subsequent estimated value according to the new value obtained every 3 minutes. In the internal model control structure, the forward model and the reverse model of the controlled system are directly added to the feedback loop. The forward model of the system is connected in parallel with the actual system, and the difference between the two is used as the feedback signal. This feedback signal is processed by the filter and controller of the forward channel. According to the nature of internal model control, the controller is directly related to the inverse of the system. The purpose of introducing the filter is to obtain the desired robustness and tracking response.

Xiao J[42] tried to use the artificial neural network model to simulate the operation process of the rectification tower. He also sought the best process conditions. The feed ratio, reflux feed ratio and product volume flow rate and catalyst volume ratio are used as the three nodes of the input layer. The conversion rate of the product and the acid-water ratio in the tower kettle are the two nodes of the output layer. He used BP algorithm combined with simulated annealing algorithm. At the same time, he used small-scale experimental data to train the network and test the training results. He interpolated or extrapolated a series of hypothetical process conditions, making sure the network predict the operating results correctly. Finally, he determined the best operation targetand found the best process through the network model condition. The results show that: based on sufficient and reliable experimental data, the neural network model can accurately predict the experimental results. Using the artificial neural network model to optimize the process conditions, satisfactory results can be obtained.

Liu J[43] proposed the use of BP neural network control technology to improve product quality. The traditional rectification tower temperature control system uses discrete PID control. Compared with traditional PID control, neural network control has many advantages. The BP neural network control method is applied to the temperature control system of the rectification separation tower. The operation results show that this control method has high control accuracy, strong adaptability and good control effect. The main raw material for industrial gas is crude propane, which contains components above $C_5$ (commonly known as green oil). This component seriously affects the quality of the product. The separation technology of the rectification tower is usually used to separate the green oil to produce crude propane with a purity greater than 95 in percent. The control of the distillation tower directly affects the product quality, output and energy consumption of the factory. Common phenomena in the chemical and thermal processes are often caused by factors such as the different speeds of exothermic and endothermic and the irreversible chemical reaction process. In the temperature control process with pure hysteresis characteristics, conventional PID control can hardly meet the requirements of control accuracy. In the classical process control theory, this kind of problem has not been solved well. Process control is difficult to describe with mathematical models. Neural networks have the ability to approximate arbitrary non-linear mappings, which can be used to establish the input and output models of the system. This paper proposes a control method for the distillation tower control system based on BP neural network and applies it to the temperature control process of the isopropanol refining tower. It satisfies the requirements of high precision and rapidity of temperature.

Zhang Y[44] selected the top temperature control system of the propylene distillation tower and the decoupling control system between the top and bottom temperatures as the main content of the research. According to problems in the process control system of the distillation column, he conducted in-depth simulation research by using neural network control. The relevance, complexity and uncertainty in the production process of the rectification tower are difficult to control. Also, the traditional control method cannot meet the requirements of production control. First, the control requirements and interference factors of the distillation column are briefly analyzed. Then the process flow of the propylene distillation column process control system is introduced, and the difficulties and problems in the control are summarized. After that, it discusses the application research of neural network in the system with difficult to determine model and time-varying problem. Aiming at the problem that it is difficult to obtain an exact mathematical model for the temperature at the top of the propylene distillation tower, a radial basis function (RBF) neural network is used to identify the output of the object. For the time-varying problem in its operation, PID controller and RBF-PID adaptive controller are designed to perform simulation experiments separately. The experimental results prove that the neural network PID controller designed has better control effect and can solve the problems in the process control system. The controller also can well solve the time-varying problem in the process control system. Finally, the application research of neural network in the system with multivariable, severe coupling and time-varying problems is discussed. Aiming at the coupling between the temperature at the top of the propylene distillation tower and the temperature at the bottom of the tower, the neural network decoupling compensation method is used for decoupling. Aiming at the time-varying problems in it, the controller adopts PID and neural network PID respectively. The simulation experiment verifies that the decoupling control using neural network method can finally obtain better control results.

Using easy-to-measure process variables (often called auxiliary variables or secondary variables), based on the mathematical relationship (soft measurement model) between these easy-to-measure process variables and the process variables to be measured that are difficult to directly measure (often called dominant variables), the measurement or estimation of the process to be measured can be realized through various mathematical calculations and estimation methods. At present, soft measurement technology has become one of the research hotspots in the field of process control.

On the basis of fully digesting and absorbing previous research results, Lu N[45] took the gas fractionation device of a refinery as the background and fully analyzed the characteristics of the system. He found a very representative propylene distillation tower according to the specific conditions of the production device. He used the soft measurement technology of artificial neural network. The real-time data of each auxiliary variable was obtained by the DCS upper computer as the training sample. BP neural network and RBF fuzzy neural network were used to model the propylene purification tower, and the predicted value was compared with the laboratory value. It is proved that the identified object model can well show the dynamic behavior of the object and has good generalization performance.

## 5 Conclusion

The propylene distillation process and the control system of the device are briefly introduced. On the basis of specific conditions of the device, analysis of the characteristics of the system and the summary of the previous research results, this paper finds out the most representative ones from problems in the propylene distillation tower.

Using predictive control, artificial neural network and other technologies, the method of using multilayer forward neural network and fuzzy neural network for nonlinear model identification is studied in view of the more complex characteristics of the nonlinear system mechanism. With the guidance of generalization theory , the problems of sample data collection and network structure determination can be solved. On the basis of theoretical analysis, the plant propylene distillation tower is dynamically identified and a prediction model is established. By selecting data to test the identification result model, the analysis results show that the identified object model can better show the dynamic behavior of the object and has good generalization performance. It lays a certain foundation for advanced control online application.

Theoretical analysis and experimental research show the superiority of neural networks in the field of control. Neural networks can approximate arbitrary nonlinear mappings with arbitrary precision, and bring a new and unconventional expression tool to the modeling of complex systems. The multiple input output structure model can be easily applied to multi-variable control systems. At the same time, because it can fuse qualitative and quantitative data, it can use the connectionist structure combined with traditional control methods and symbolic artificial intelligence.